\documentclass[12pt,preprint]{aastex}

\shorttitle{Morphological Dependence of MIR Properties}
\shortauthors{Li et al.}

\newcommand{\NIIHa}{\ion{N}{2}/H$\alpha$}
\newcommand{\OIIIHb}{\ion{O}{3}/H$\beta$}

\begin{document}

\title{Morphological Dependence of MIR Properties \\
of SDSS Galaxies in the {\it Spitzer} SWIRE Survey}

\author{Hai-Ning Li\altaffilmark{1,2},
Hong Wu\altaffilmark{1,3}, Chen Cao\altaffilmark{1,2}, Yi-Nan
Zhu\altaffilmark{1,2}}

\altaffiltext{1}{National Astronomical Observatories, Chinese
Academy of Sciences,
        A20 Datun Road, Beijing 100012, P.R.China;
        lhn@bao.ac.cn; hwu@bao.ac.cn}
\altaffiltext{2}{Graduate School of Chinese of Academy of Sciences,
        Beijing 100039, P.R.China}
\altaffiltext{3}{Visiting Scholar, Institute for Astronomy, University of Hawaii,
        2680 Woodlawn Drive, Honolulu, HI 96822}
\received{\data}
\accepted{}

\begin{abstract}

We explore the correlation between morphological types and
mid-infrared (MIR) properties of an optically flux-limited sample of
154 galaxies from the Forth Data Release (DR4) of Sloan Digital Sky
Survey (SDSS), cross-correlated with {\it Spitzer} SWIRE ({\it
Spitzer} Wide-Area InfraRed Extragalactic Survey) fields of
ELAIS-N1, ELAIS-N2 and Lockman Hole. Aperture photometry is
performed on the SDSS and {\it Spitzer} images to obtain optical and
MIR properties. The morphological classifications are given based on
both visual inspection and bulge-disk decomposition on SDSS $g-$ and
$r-$band images. The average bulge-to-total ratio ($B/T$) is a
smooth function over different morphological types. Both the $8\mu
m$(dust) and $24\mu m$(dust) luminosities and their relative
luminosity ratios to $3.6\mu m$ (MIR dust-to-star ratios) present
obvious correlations with both the Hubble $T$ type and $B/T$. The
early-type galaxies notably differ from the late-types in the MIR
properties, especially in the MIR dust-to-star ratios. It is
suggested that the MIR dust-to-star ratio of either $\nu
L_{\nu}[8\mu m(dust)]/\nu L_{\nu}[3.6\mu m]$ or $\nu L_{\nu}[24\mu
m(dust)]/\nu L_{\nu}[3.6\mu m]$ is an effective way to separate the
early-type galaxies from the late-type ones. Based on the tight
correlation between the stellar mass and the $3.6\mu m$ luminosity,
we have derived a formula to calculate the stellar mass from the
latter. We have also investigated the MIR properties of both edge-on
galaxies and barred galaxies in our sample. Since they present
similar MIR properties to the other sample galaxies, they do not
influence the MIR properties obtained for the entire sample.

\end{abstract}



\section{INTRODUCTION}\label{section:introduction}

Ever since Hubble's famous paper outlined his classification system
\citep{Hubble1926,Hubble1936}, morphological classification in
conjunction with physical measurement has become an important tool
in extragalactic astrophysics. A number of quantitative classifiers
have been developed or extended over the years to probe the
structure of galaxies. There are parametric classifiers like radial
multi-Gaussian deconvolution \citep{Bendinelli1991}, bulge-disk
decomposition \citep{Byun1995}, etc., and the nonparametric ones such as
the C-A system \citep{Abraham1994,Abraham1996}, artificial neural
nets trained from visual classification sets \citep{Odewahn1996},
Gini Coefficient \citep{Lotz2004} and so on. As one of the main
quantitative criteria and a function of the Hubble classification,
the bulge-disk decomposition has now been widely used as an effective
method to examine galaxy structures and morphological properties,
\citep[e.g.,][etc]{deJong1996,Baggett1998,Tasca2005,Allen2006}.

The variations of galaxy physical properties with morphology and
environment are crucial in our understanding of the evolution of
galaxies \citep{Kennicutt1998,Brinchmann2004}. Numbers of properties
such as the integrated birthrate variation \citep{Sandage1986}, the
optical and infrared photometric properties
\citep{Boselli2001,Shimasaku2001,Popescu2002a,Davoodi2006}, the star
formation properties of galaxies in clusters \citep{Yuan2005}, the
circumnuclear H$\alpha$ luminosity and bar structures
\citep{Shi2006}, etc, which have been investigated present
regularity or correlation along different morphological types.
Through substantial former investigations, it is known that as a
result of different stellar populations and the amount of dust and
gas for the environment of star-forming, the early-type galaxies
(ellipticals and lenticulars) exhibit rather different properties
compared with the late-type ones (spirals, irregulars and so on).

As we know, averagely one third of the total luminosity from normal
galaxies is absorbed and re-radiated by dust
\citep{Mathis1990,Popescu2002b}, and even higher fraction from
galaxies with the most active star-forming activity \citep[e.g.
luminous infrared galaxies (LIGs),][]{Sanders1996,Wang2006}. Both
the Infrared Space Observatory \citep[ ISO - ][]{Kessler1996} and
{\it Spitzer} Space Telescope \citep{Werner2004} continued to
explore the importance of dust. Studies of dusty starburst galaxies
\citep{Poggianti2000,Flores2004} have shown that, most of the
activities (e.g., star-formation and/or AGN emission) in these
galaxies are hidden by dust, and the bolometric luminosities of the
active systems are mostly emitted in the infrared. This suggests
that the infrared emission is a sensitive tracer of the young
stellar population and star formation rates (SFRs), and suffers
weaker extinction. The MIR dust emissions mainly consist of
polycyclic aromatic hydrocarbons (PAHs) emission and the continuum
emission feature of warm dust component. Both of these two
components have been investigated as reliable measures of the SFRs
of galaxies as a whole \citep{Peeters2004,Wu2005a}. Whereby studies
in the MIR properties will certainly improve our understanding of
the galaxies which have been well studied in the optical bands, and
give an insight into the details of their star formation histories.

This work tries to explore the relationships between the morphology
and the MIR properties, for a flux-limited sample of normal galaxies
which were selected from the galaxies of SDSS-DR4
\citep{Adelman-McCarthy2006} cross-correlated with the {\it Spitzer}
SWIRE \citep{Lonsdale2003}. Considering that the {\it Spitzer} IRAC
(Fazio et al. 2004) $8\mu m$ band covers the strongest PAH feature
($7.7\mu m$), and the MIPS (Rieke et al. 2004) $24\mu m$ band covers
the continuum emission of very small grains (VSGs) free of PAH
features, we tried to employ the IRAC bands and MIPS $24\mu m$ band
to investigate the MIR dust properties of entire galaxies. We
performed elliptical aperture photometric analysis in both the
optical and MIR bands. The $8\mu m$ dust and $24\mu m$ dust
luminosity and their dust-to-star ratios are used to quantitatively
investigate the MIR properties of our sample galaxies. Two ways to
classify the morphologies of the sample galaxies were adopted: the
$T$ type of the revised Hubble classification system
\citep{deVaucouleurs1976} by visual inspection, and the quantitative
parameter of $B/T$ by the bulge-disk decomposition.

In \S 2, we describe the infrared and optical data, the sample
construction, and the elliptical aperture photometry. The
morphological classifications are presented in \S 3. In  \S 4, we
analyze the optical and MIR colors, and give the statistical results
of the MIR dust properties against galaxy morphological types and
the corresponding discussions. The conclusions are presented in \S
5. Throughout this paper, we assume a Hubble constant $H_{0} = 70
{\rm km s^{-1} Mpc^{-1}}$ and $\Omega_M=0.3$, $\Omega_\Lambda=0.7$
in calculating the distance and the luminosity.


\section{DATA AND DATA REDUCTION}\label{sec:data.process}

\subsection{The Sample}\label{subsec:sample}

We used the IRAC $3.6, 4.5, 5.8, 8.0\mu m$ and the MIPS $24\mu m$
images from the northern SWIRE fields of Lockman Hole, ELAIS-N1, and
ELAIS-N2. The BCD (Basic Calibrated Data) images of the four IRAC
bands were obtained from {\it Spitzer} Science Center, which include
flat-field corrections, dark subtraction, linearity and flux
calibrations \citep{Fazio2004}. The IRAC images (in all four bands)
were mosaiced from the BCD images after pointing refinement,
distortion correction and cosmic-ray removal with the final pixel
scale of $0.6\arcsec$ as described by \citet{Huang2004} and
\citet{Wu2005a}; likewise the MIPS $24\mu m$ images were mosaiced in
the similar way with the final pixel scale of $1.225\arcsec$
\citep{Cao2007,Wen2007}. Sources detected by SExtractor
\citep{Bertin1996} in IRAC four bands and MIPS $24\mu m$ were
matched with the Two Micron All Sky Survey (2MASS) sources to
achieve astrometric uncertainties of around $0.1\arcsec$.

The SDSS data provide full coverage of the SWIRE fields of Lockman
and ELAIS-N2 but cover only one third of ELAIS-N1 field. The $ugriz$
corrected frames of spectroscopically observed galaxies were taken
from the SDSS-DR4. The pixel scale of SDSS images is $0.4\arcsec$
\citep{Stoughton2002}. The SDSS-DR4 $ugriz$ spectrophotometric
catalogue was cross-correlated with SWIRE MIR catalogue measured by
SExtractor \citep{Bertin1996} by a radius of $2\arcsec$. The total
survey area of the three northern SWIRE fields is $\sim$24
deg$^{2}$, of which the overlap with SDSS is $\sim$15 deg$^{2}$.
To obtain the reliable morphological classification
\citep{Fukugita2004}, 163 bright galaxies with Petrosian magnitude
$r\leqslant15.9$ were selected. Only four SDSS galaxies were not
matched with SWIRE MIR sources by 2$\arcsec$, since the SDSS fiber
observations mistakenly pointed to the off-nucleus regions rather
than the nuclear regions of galaxies. This provides a preliminary
sample of 159 galaxies.

\begin{figure}[t]
\epsscale{1.0} \plotone{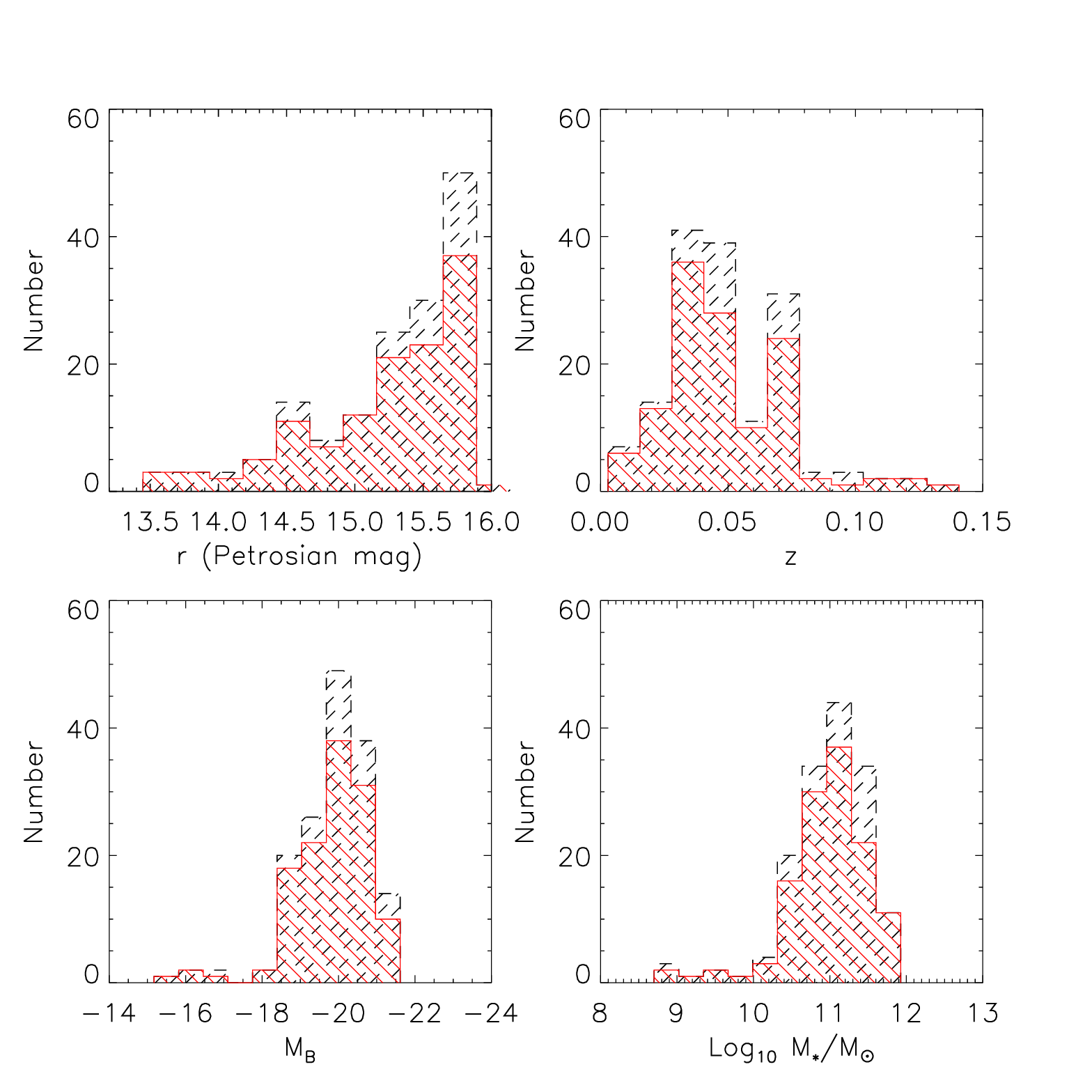} \figcaption{Distribution of sample
galaxies' SDSS $r-$band Petrosian magnitudes, the redshifts, B-band
absolute magnitudes and the stellar mass (deduced from
\S~\ref{subsec:I1.mass}). Black histograms represent distribution of
all the 154 sample galaxies, while the red for the 125 sample
galaxies which have photometric information in all MIR bands.
\label{fig:properties} }
\end{figure}

We excluded further five galaxies in our magnitude limited sample of
159 galaxies, because they failed in either the sky-background
subtraction or the bulge-disk decomposition, or were severely
contaminated by nearby bright stars. This led to a SDSS $r-$band
flux-limited final sample of 154 galaxies. 142 of this sample have
been imaged by IRAC four bands and 137 have been imaged by MIPS
24$\mu$m band. Finally, a sub-sample of 125 galaxies has images in
all five MIR bands, and is used for further statistical discussion
in \S\ref{subsec:statistics}. All of theses objects are local, with
redshift less than 0.13. In our sample, six galaxies were with the
absolute B magnitude fainter than -18, and thus classified as dwarf
galaxies \citep{Mateo1998}. Here, the B magnitude can be obtained
from the SDSS $g-$ and $r-$ magnitudes \citep{Smith2002}. The
distributions of SDSS $r-$band Petrosian magnitudes, the redshift,
and B-band absolute magnitudes for the 154 sample galaxies as well
as those for the 125 sub-sample galaxies are plotted in
Figure~\ref{fig:properties}. All these distributions show that the
sub-sample can well represent the flux-limited sample.

\begin{figure}[t]
\epsscale{1.0} \plotone{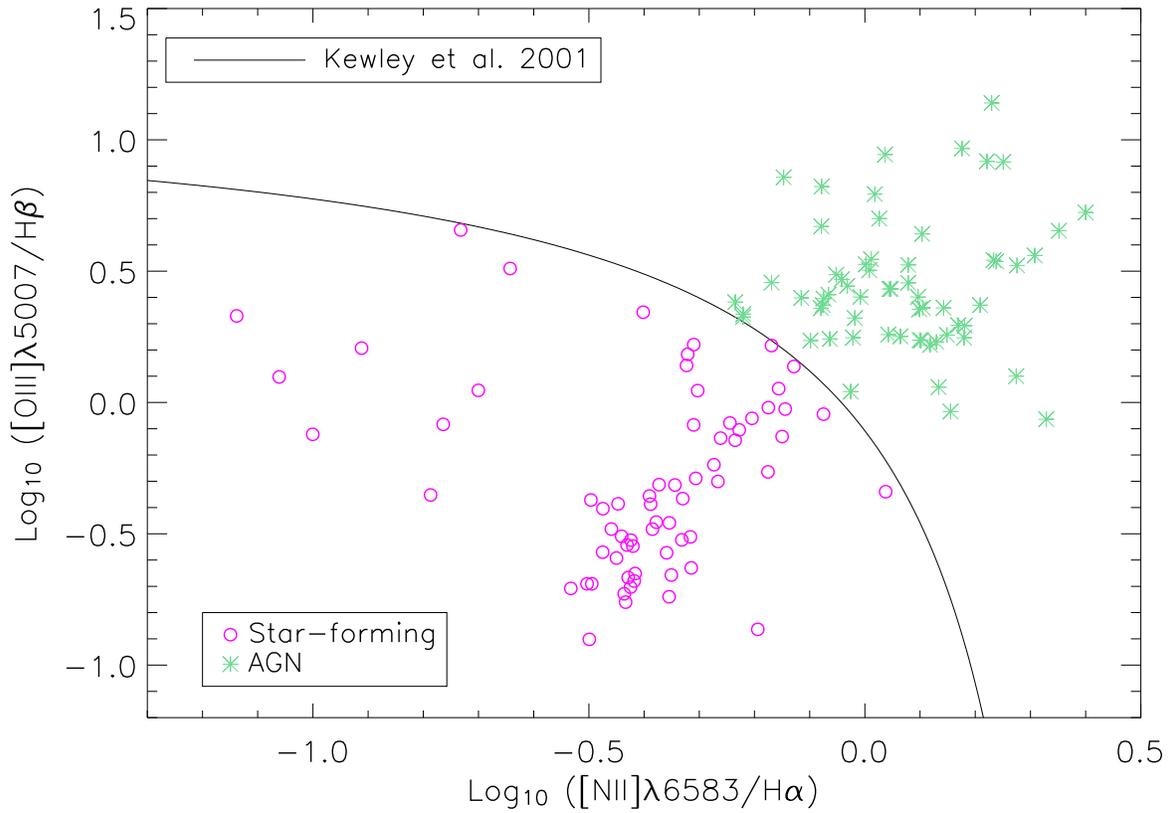} \figcaption{BPT diagram plotted for
134 sources with emission-line detections.
    The solid curve illustrates the criteria based on the model of Kewley et al.(2001).
    Star-forming galaxies are labeled with magenta circles and AGNs with green asterisks.
\label{fig:BPT}
}
\end{figure}

Among the 154 sample galaxies, 133 with emission line detections
could be classified with the traditional
$\log$(\OIIIHb)-$\log$(\NIIHa) diagnostic diagram
\citep{Baldwin1981,Veilleux1987}. The emission line fluxes were
derived from the  catalogue of \citet{Tremonti2004}. The criteria
given by \citet{Kewley2001} was adopted to distinguish the potential
star-forming galaxies from AGNs, as is shown in
Figure~\ref{fig:BPT}.

\subsection{Photometry}\label{subsec:photometry}
\begin{figure}[t]
\epsscale{0.9} \plotone{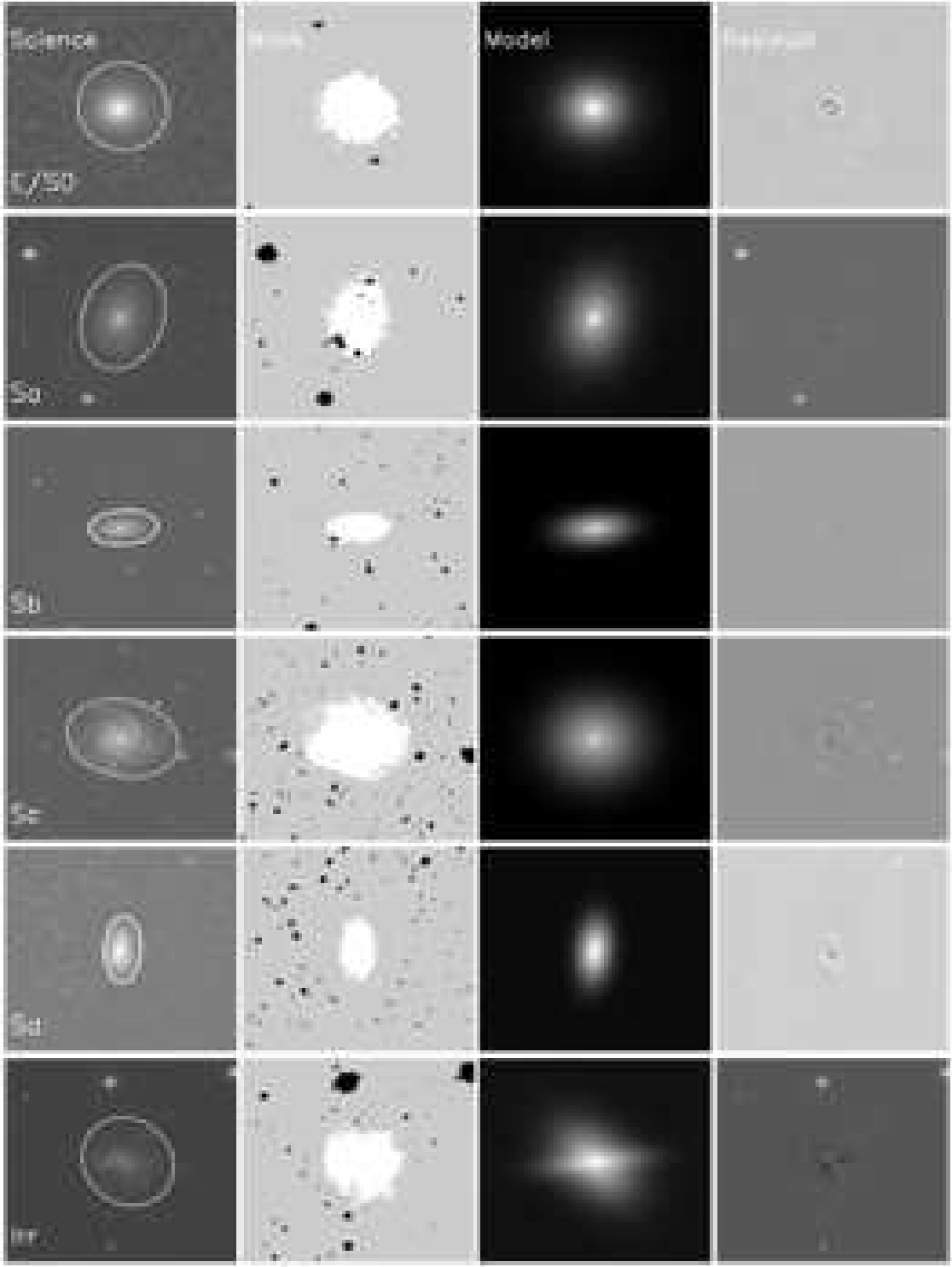} \figcaption{\footnotesize Examples
of the science, mask, model and residual SDSS $r-$band images
    for our sample, from top to bottom for galaxies corresponding to our morphological types:
    E/S0, Sa, Sb, Sc, Sd, and Irr. Elliptical isophotes for aperture photometry are overlaid
    in the first column.
\label{fig:sample}
}
\end{figure}

To obtain the accurate photometry, sky background fitting and
subtraction were done \citep{Zheng1999,Wu2002,Wu2005b} on both the
SDSS corrected frames and the {\it Spitzer} images. All objects
detected by SExtractor were masked to generate a background-only
image, and the fitting sky-background was then subtracted.
Photometry was performed on the background-subtracted images by IRAF
task ELLIPSE \citep{Jedrzejewski1987}. To embrace almost all the
flux of these extended sources in the different bands, an elliptical
isophote with the B-band surface brightness of 26 mag $arcsec^{-2}$
was adopted as the photometric aperture, based on the SDSS $g-$band
images. With such elliptical apertures, the total fluxes were
measured in all the wavelengths including the SDSS $ugriz$, the IRAC
four bands and the MIPS $24\mu m$ band (see sample apertures marked
in Figure~\ref{fig:sample}) with IRAF task POLYPHOT. Note that the
point source functions (PSFs) of MIPS $24\mu m$ are rather extended,
therefore further aperture corrections were performed on this band.
For objects with the equivalent radius of the elliptical apertures
smaller than $15\arcsec$, aperture corrections were applied to
calibrate the integrated flux to an equivalent radius of
$15\arcsec$. The photometric accuracies of these different bands are
quite small, less than 0.03 mag on average. Flux calibration
accuracies of the IRAC four bands \citep{Fazio2004} and MIPS $24\mu
m$ band \citep{Rieke2004} are less than $10\%$. The final errors
include both the above errors.

The Galactic extinction in each SDSS filter from SDSS-DR4 catalogue
was adopted, and then intrinsic extinction was derived from the
Balmer decrement \citep{Calzetti2001}. The photometric
$K$-correction was calculated using the method of \citet[][{\rm
Kcorrect V4-1-4}]{Blanton2003}. No extinction correction has been
performed on photometric results in the MIR bands since extinction
effect is rather negligible in the infrared compared with optical
wavelength. Due to the fact that our sample are all low redshift
galaxies and, as of this work, there are no reliable $K$-correction
for these MIR spectral ranges available yet, we applied no
$K$-correction to the MIR photometry. All the measurements were
converted to AB magnitudes \citep{Oke1983}.

Although PAH and VSG emissions dominate $8\mu m$ and $24\mu m$ bands
for the majority of our sample, there is still a stellar continuum
in these bands, especially for the early-type galaxies. Herewith a
subtraction of the stellar contribution using the $3.6\mu m$
luminosity was adopted, with a scale factor of 0.232 for the $8\mu
m$ band and 0.032 for the $24\mu m$ band \citep{Helou2004}.
Hereafter, we denoted the 8$\mu m$(dust) and 24$\mu m$(dust)
representing the dust emissions after subtracting the stellar
contribution \citep{Wu2005a}.


\section{MORPHOLOGICAL CLASSIFICATION}\label{sec:mor.class}
We conducted the morphological classification by two methods:
visual inspection and bulge-disk decomposition with GIM2D
\citep[Galaxy IMage 2D:][]{Simard1998,Simard2002}.

\subsection{GIM2D Fitting}\label{subsec:gim2d.fit}
GIM2D is a two-dimensional photometric decomposition fitting
algorithm which fits each image to a superposition of an bulge
component with a S\'{e}rsic profile, and a disk component with an
exponential profile \citep{Simard1998,Simard2002}. GIM2D was
employed to obtain the structural parameters of galaxies in our
sample. The bulge component of the model is a profile of S\'{e}rsic
form \citep{Sersic1968}:
\begin{equation}
    \Sigma(r) = \Sigma_e\cdot exp\{-b[(r/r_e)^{1/n}-1]\}
\label{eqn:Sersic.profile}
\end{equation}
where $\Sigma(r)$ is the surface brightness at a  radius $r$ and
$\Sigma_e$ is the characteristic value (i.e. the effective surface
brightness), defined as the brightness at the effective radius
$r_e$. Parameter $b$ is related to the S\'{e}rsic index $n$ and
chosen equal to $1.9992n-0.3271$ so that $r_e$ remains the projected
radius enclosing half of the light in this component
\citep{Ciotti1991}.

The disk component is an exponential profile of the form:
\begin{equation}
    \Sigma(r) = \Sigma_0\cdot exp(-r/r_d)
\label{eqn:Exponential.profile}
\end{equation}
where $\Sigma_0$ is the central surface brightness,
and $r_d$ is the disk scale length.

Decomposition was performed based on this model, with a Gaussian
PSF. A total of twelve parameters were adjusted in fitting the
galaxy image and retrieved as output from our decomposition: the
total luminosity $L$, the bulge fraction $B/T$, the bulge effective
radius $r_e$, the bulge ellipticity $e$, the bulge position angle
$\phi_b$, the disk scale length $r_d$, the disk inclination angle
$i$, the disk position angle $\phi_d$, the centroiding offsets $dx$
and $dy$, the S$\acute{e}$rsic index $n$, and the residual sky
background level $db$. The $B/T$ which is defined as the fraction of
the total flux in the bulge component has been extracted as a
quantitative indicator of morphology. $B/T=1$ corresponds to a pure
bulge, while $B/T=0$ to a pure disk. With all the objects detected
by SExtractor flagged except the galaxy of interest, the $g-$band
mask images were adopted not only in $g-$band but also in $r-$band,
throughout the decomposition fitting with GIM2D.
Figure~\ref{fig:sample} shows example (science, mask, model, and
residual) images of different morphological types in GIM2D fitting.
The returned $\chi ^2$ values of $g-$ and $r-$band fitting have the
mean values of 1.14 and 1.18 with deviations of 0.26 and 0.33
respectively, representing rather convincing fitting results.

\begin{figure}[t]
\epsscale{1.0} \plotone{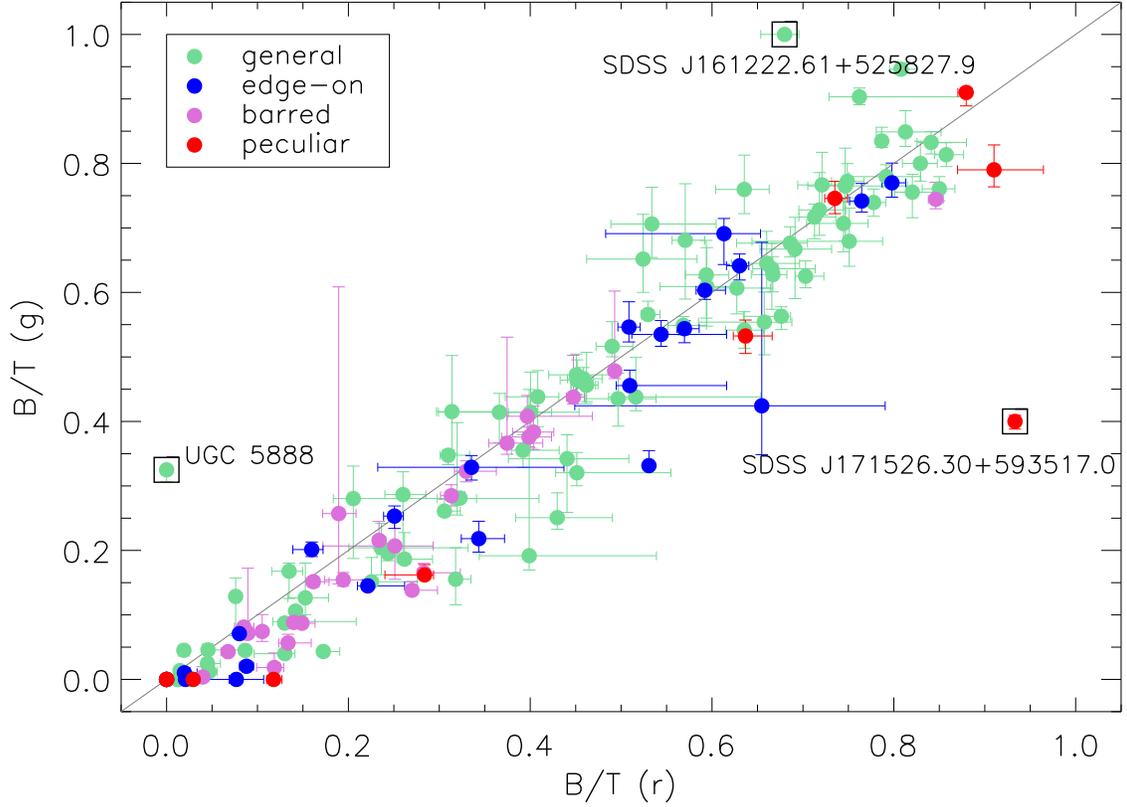} \figcaption{The bulge-to-total
ratios $B/T$ in the SDSS $r-$ and $g-$band are plotted
    against each other for the 154 galaxies modeled
    in both bands by fitting a S\'{e}rsic profile to the bulge and an
    exponential to the disk. Error-bar representing $99\%$ confidence are plotted.
    Note that different morphological types are plotted,
    with red filled circles for peculiar galaxies, green for general,
    blue and orchid representing edge-on and barred types respectively.
    The diagonal is plotted for comparison.
\label{fig:gim2d} }
\end{figure}

To check the results obtained, we compared the parameters $B/T$ of
the 154 galaxies estimated from $g-$ and $r-$band in
Figure~\ref{fig:gim2d}. The $B/T$ values obtained from both bands
agree well with small amount of deviation. Among the three most
deviated sources, SDSS J161222.61+525827.9 is an early-type galaxy
with highly centralized surface brightness distribution in bluer
band. SDSS J151723.30+593517.0 is a peculiar galaxy and UGC5888 is
an irregular. We adopted $B/T$ values derived from $r-$band images
throughout the following investigation since the decomposition
results in the two SDSS bands hold quite good agreement.

\subsection{Visual Classification}\label{subsec:visual.class}

The morphological types of revised Hubble sequence
\citep{deVaucouleurs1964} of our sample galaxies were classified by
visual inspection based on features like bulge ratios, the presence
of spiral arms and/or bars, signs of interaction, multiple nuclei
etc. All galaxies in our sample were classified into six
morphological classes: $T=0$(E or S0), 1(Sa), 3(Sb), 5(Sc), 7(Sd),
and 9(Irr). Notice that we assigned an additional class 10
corresponding to galaxies with peculiar morphology possibly related
to galactic interactions, mergers, etc. We performed the visual
classification in both SDSS $g-$ and $r-$band images. The
classifications done independently by the four of us, agree in over
90\% of the sample. The classifications were verified for 36 of
those galaxies whose morphological types were found in the NASA/IPAC
Extragalactic Database (NED)
\footnote{See http://nedwww.ipac.caltech.edu/index.html}
and were found to be accurate to $\Delta T=\pm1$.

Considering the possible effects that inclinations and bar
structures may cause, we further divided sample galaxies into 3
types: barred galaxies, edge-on galaxies, and the rest defined as
general galaxies for further consideration. Note that when defining
edge-on galaxies, we adopted a standard of GIM2D fitted disk
inclination angle $i > 70$. Table~\ref{tab:mor.class} shows the
numbers and corresponding fractions of different morphological types
in our sample. In Table~\ref{tab:mor.spitzer}, we present the
numbers of galaxies observed by different MIR bands in each
morphological type. It can be drawn from Table 2 that the
morphological fractions E/S0: S(Sa-Sd): Irr of optical-$8\mu m$ and
optical-$24\mu m$ samples are 0.47: 0.51: 0.014 and 0.44: 0.54:
0.016 respectively. Both are consistant with E(E/S0-S0): S(S0a-Sdm):
Im of 0.40: 0.57: 0.014 obtained from an optically selected sample
of SDSS galaxies \citep{Fukugita2007}.

\subsection{Comparison}\label{subsec: gim2d.compare}
\begin{figure}[t]
\epsscale{1.0} \plotone{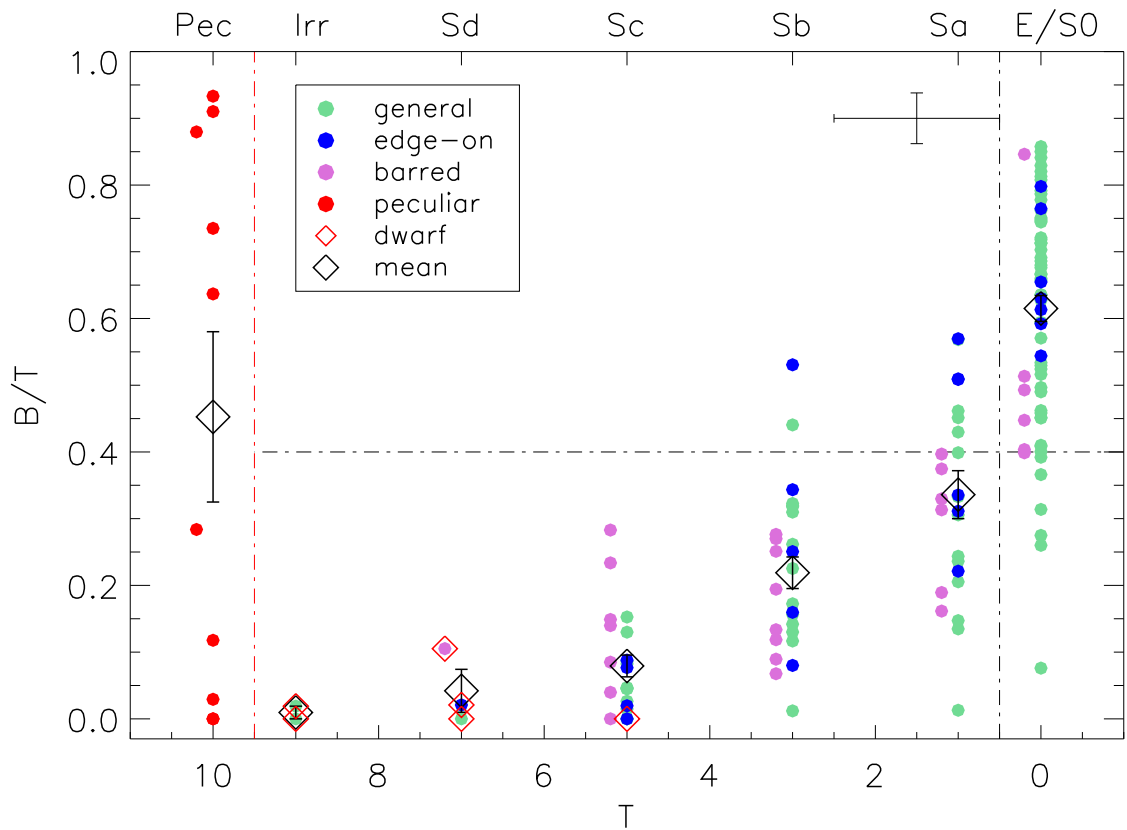} \figcaption{This plot exhibits the
correlation between the bulge-to-total ratio $B/T$ and $T$-type.
    Different colors of filled circles are defined as in Figure~\ref{fig:gim2d}.
    The mean value of each type(including barred, edge-on, and general types)
    are overlaid in black diamonds, with standard deviation shown.
    Dwarf galaxies are overlaid with red diamonds.
    Divisions of normal galaxies with early-type and late-type are plotted, by
    $T$ and $B/T = 0.4$, with peculiar galaxies excluded.
    The typical errors of sample galaxies are plotted in the upper right corner.
\label{fig:T.Bratio} }
\end{figure}

A comparison between Hubble $T$ type and the bulge-to-total ratio
$B/T$ has been carried out, as is shown in
Figure~\ref{fig:T.Bratio}. Except for the peculiar galaxies which
exhibit a diversity of $B/T$, there is a smooth inclination in $B/T$
along Hubble sequence from $T=9$ to $T=0$, i.e., for normal
galaxies, except the peculiar, the late-types exhibit lower $B/T$
than the early-type ones. Hence $B/T$ does act as a reliable measure
of morphological types for normal galaxies in our sample. In order
to statistically examine the properties of our sample, galaxies have
been divided into two morphological types: the early-type (E/S0)
with T=0, and the late-type (Sa, Sb, Sc, Sd, and Irr) with T from 1
to 9. Consequently, according to \citet{Simien1986}, such a
classification roughly corresponds to the division of $B/T=0.4$. 128
out of the 144 normal sample objects coincide with their
morphological types classified with $T$. The divisions are shown in
Figure~\ref{fig:T.Bratio}. Since as a whole, divisions with the two
methods agree with each other, we have adopted T type to divide the
sample galaxies into either the early-type or the late-type in the
following statistics.


\section{RESULT AND DISCUSSION}\label{sec:result}
\subsection{Color-Color Diagram}
\begin{figure}[t]
\epsscale{1.0} \plotone{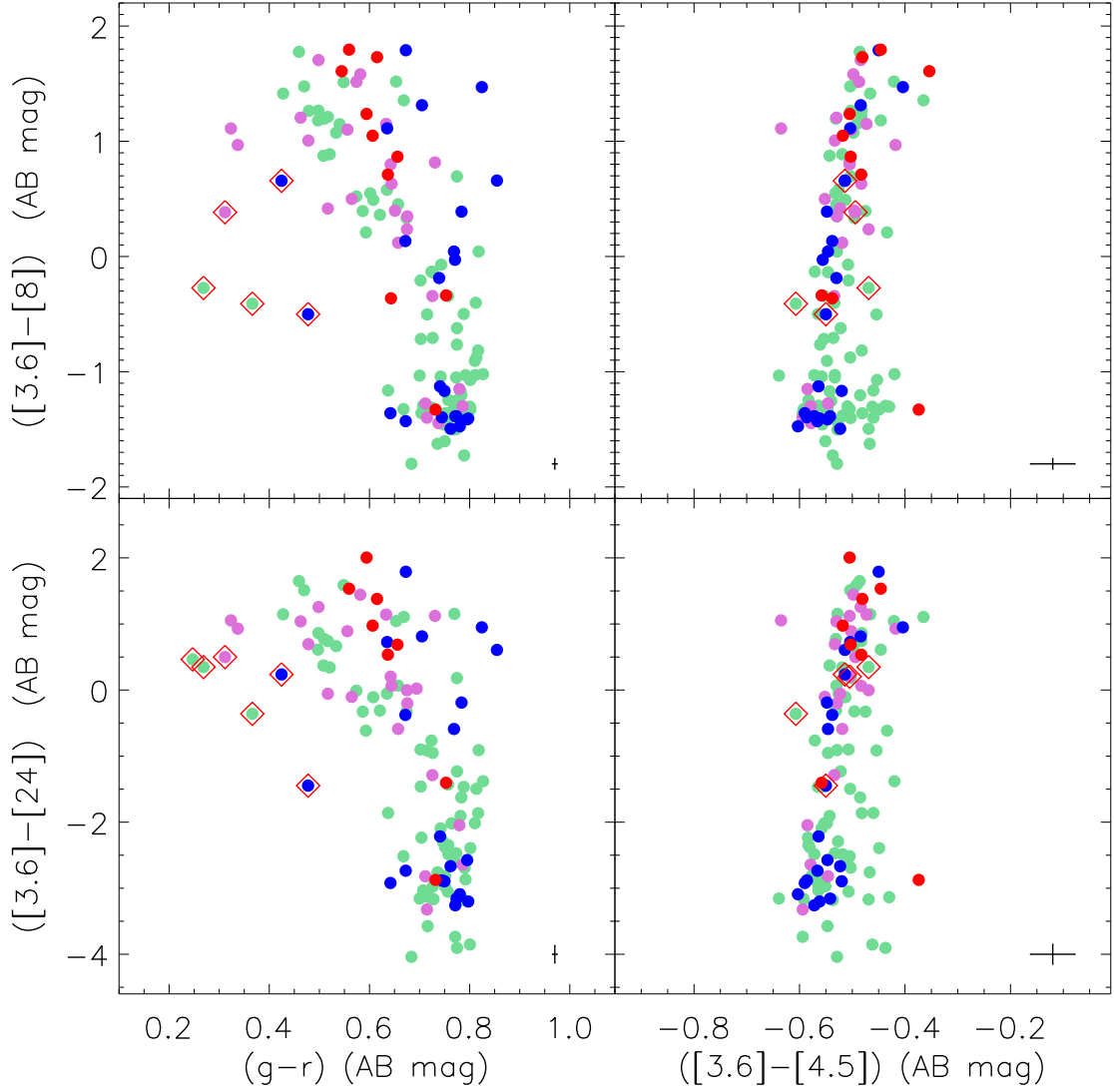} \figcaption{Color-color diagrams of
our sample.
    The left panels show the derredened g-r color against MIR colors,
    while the right for MIR color-color diagram.
    Symbols are defined as in Figure~\ref{fig:T.Bratio}.
    The typical errors are plotted in the lower right corners.
\label{fig:color} }
\end{figure}

The optical-MIR color-color diagrams are shown in the left panels of
Figure~\ref{fig:color}, and there are anti-correlations between
optical and MIR colors, which is consistent with the result of
\citet{Hogg2005}. Such a trend can be explained rather naturally:
the optically blue color is related to the recent star formation,
indicating the existence of notable amount of dust, and the MIR dust
emissions can be shown in both $8\mu m$ and $24\mu m$ bands. On the
other hand, optically red galaxies are always too old to contain
plenty of dust, and hence present weak MIR dust emissions.

The right panels of Figure~\ref{fig:color} show the relationship
between MIR colors. It can be detected that [3.6]-[4.5] colors in
our sample are quite blue, with an average of around -0.52,
indicating that there is no extremely active galactic nuclei (QSO
etc.) in our sample, because both the emissions of $3.6\mu m$ and
$4.5\mu m$ of normal galaxies are dominated by decreasing stellar
continuum of the old stellar population. Yet these two bands of very
active galaxies like quasars are dominated by the power law spectra
of quasars and consequently should exhibit redder [3.6]-[4.5]
colors. Both colors of [3.6]-[8] and [3.6]-[24] can roughly
characterize the relative strengths of the MIR dust emissions of
PAHs and VSGs. The stronger the dust emissions, the redder the
colors. Therefore, it is not surprised to find that most of the
peculiar galaxies of our sample at the redder part of these panels
because they contain large amount of dust for their violent star
formation \citep{Sanders1996}.

\subsection{Estimation of the Stellar Mass}\label{subsec:I1.mass}
The luminosity of $3.6\mu m$ band is often treated as a tracer of
stellar component \citep[][etc]{Wu2005a,Davoodi2006} as well as a
test of validity of mass determination \citep{Hancock2007}.
Furthermore \citet{Hancock2007} has compared the mass of clumps in
Arp 82 derived from {\it R} band fluxes and broadband colors against
luminosities of $3.6\mu m$, and has found strong correlation between
them. Hereby the relationship between the stellar mass and the
$3.6\mu m$ luminosity of our sample galaxies is examined. Based on
the optical photometries as described in \S~\ref{subsec:photometry},
we calculated the stellar mass of our sample, following
\citet{Bell2003}:
\begin{equation}
    \log(\frac{M_{\star}}{M_{\odot}}) = -0.4\times(M_{r,AB}-4.67) + [a_{r}+b_{r}\times (g-r)_{AB}+0.15]
\label{eqn:mass.Bell}
\end{equation}
where $M_{r,AB}$ is the $r-$band absolute magnitude, $(g-r)_{AB}$ is
the rest-frame color in the AB magnitude system. The coefficients
$a_r$ and $b_r$ come from Table 7 of \citet{Bell2003}. A
\citet{Salpeter1955} stellar mass initial function (IMF) has been
adopted with $\alpha = 2.35$ and $0.1M_{\odot} < M < 100M_{\odot}$.
The distribution of the stellar mass for our sample galaxies is
presented in Figure~\ref{fig:properties}, in a range between $10^{9}
M_{\odot}$ and $10^{12} M_{\odot}$, with the average mass around
$10^{11} M_{\odot}$, as intermediate mass galaxies.

\begin{figure}[t]
\epsscale{0.7}
\plotone{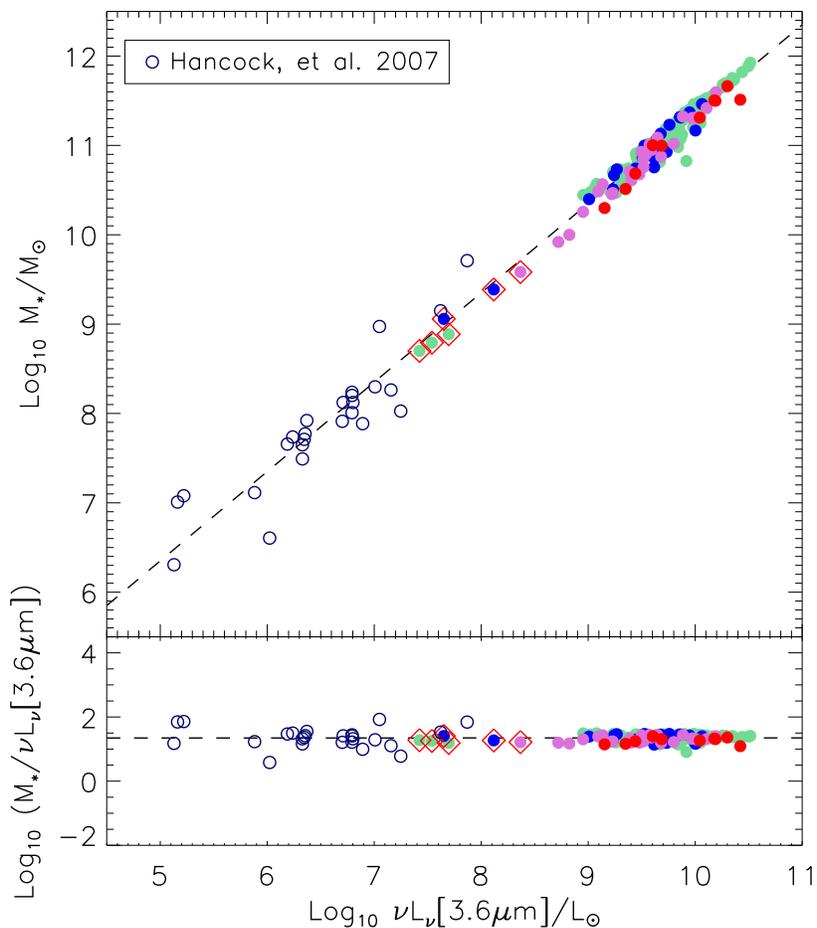}
\figcaption{Upper panel: the $3.6\mu m$ luminosity vs. the stellar mass of our sample.
    Lower panel: the ratio between stellar mass and the $3.6\mu m$ luminosity vs. the $3.6\mu m$ luminosity.
    Symbols are the same defined as in Figure~\ref{fig:T.Bratio}.
    Open circles are from \citet{Hancock2007}, corresponding to mass clumps in Arp 82.
    The typical random errors of $\log \nu L_{\nu}[3.6\mu m]$ luminosity and
    $\log M_{\star}$ of these 169 data points are 0.09 and 0.01 respectively.
\label{fig:I1.mass} }
\end{figure}

A tight correlation of the stellar mass against the luminosity of
$3.6\mu m$ is detected for our sample, as is shown in
Figure~\ref{fig:I1.mass}. Our result is consistent with that found
by \citet{Hancock2007} but with smaller scatters, thus it confirms
the capability of the $3.6\mu m$ luminosity as a measure of the
stellar mass of galaxies. Based on our 145 galaxies and 24 clumps in
Arp 82 \citep{Hancock2007}, we fit the relation between the stellar
mass and the $3.6\mu m$ luminosity, and obtain a nearly linear
correlation:
\begin{equation}
    \log(\frac{M_{\star}}{M_{\odot}}) = (1.34\pm0.09) + \\
            (1.00\pm0.01)\times \log(\frac{\nu L_{\nu}[3.6\mu m]}{L_{\odot}})
\label{eqn:I1.mass}
\end{equation}

As is pointed out by \citet{Charlot1996} and \citet{Madau1998}, the
mass-to-infrared light ratio is relatively insensitive to the star
formation history, and remains very close to unity, independent of
either galaxy colors or Hubble types. This relation is probably due
to the fact that in the near-infrared, older stellar populations may
dominate both the galaxy luminosities and the stellar masses.
Therefore such a correlation provides a proxy way to estimate the
stellar mass of galaxies directly from the integrated $3.6\mu m$
luminosity.

\subsection{MIR Properties and Morphology}\label{subsec:MIR.mor}
\begin{figure}
\epsscale{1.1}
\plotone{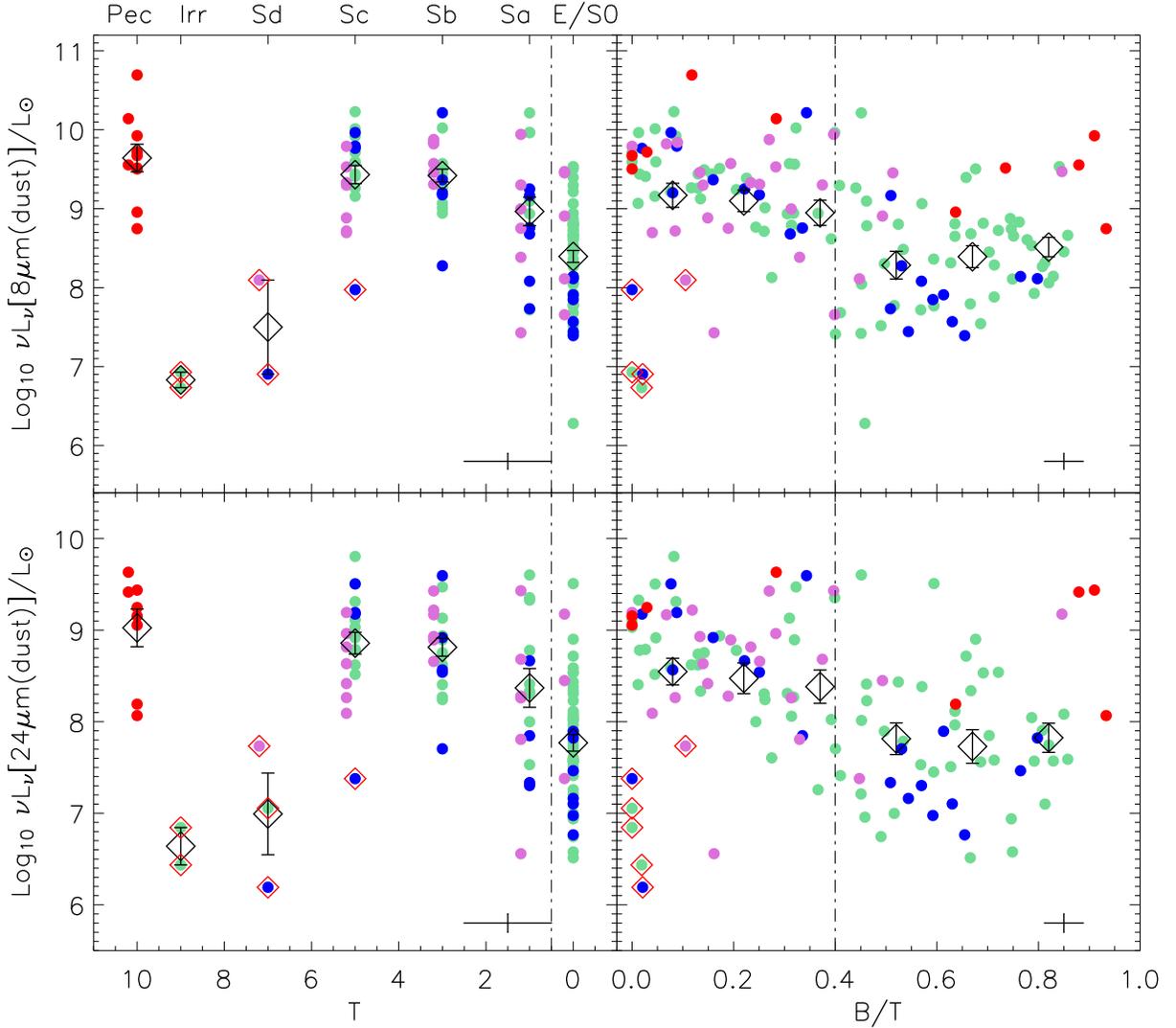}
\figcaption{Distributions of photometric MIR luminosities for our sample,
    subdivided by different morphological types.
    Left panels: (upper)$8\mu m$ dust luminosity and (lower) $24\mu m$ dust luminosity vs. $T$;
    Right panels: (upper)$8\mu m$ dust luminosity and (lower) $24\mu m$ dust luminosity vs. $B/T$.
    Symbols represent the same definitions as in Figure~\ref{fig:T.Bratio}.
    The dot-dashed line represents the divisions of early-type and late-type.
    The typical errors are plotted in the lower right corners.
\label{fig:mor.lum} }
\end{figure}

Figure~\ref{fig:mor.lum} shows the $8\mu m$ and $24\mu m$ dust
luminosities as the function of different morphological types in our
sample. For normal galaxies, except the dwarfs, along either the
Hubble $T$ types or the $B/T$ ratios, there exist obvious
declinations of the MIR luminosities from the late-type to the
early-type galaxies, especially showing a steep change around the
division of the late-type and the early-type galaxies. All the
peculiar galaxies show relatively higher MIR luminosities,
independent of their $B/T$ ratios. The phenomenon could be
attributed to the fact that early-type galaxies are dominated by
older population and are deficient of dust, resulting in lower MIR
dust luminosities, while the late-type contain larger amount of
young stars and more dust thus present stronger MIR emissions.
Peculiar galaxies which are undergoing strong star forming
activities contain great amount of dust and thereby show averagely
high MIR luminosities. As \citet{Wu2005a} has pointed out, both the
$8\mu m$ and $24\mu m$ dust luminosities can be used as measures of
SFRs of entire galaxies; therefore, the correlations between the MIR
dust luminosities and morphological types reflect a consequent
relationship between the galactic SFRs and morphological types. This
also confirms the previous results of \citet{Sandage1986},
\citet{Kennicutt1998}, etc. As for the six dwarf galaxies, because
they all have low mass of around $10^9 M_{\odot}$ (from the previous
mass determination), so they contain less dust and thus show lower
MIR dust emissions \citep{Hogg2005}.

\begin{figure}
\epsscale{1.1}
\plotone{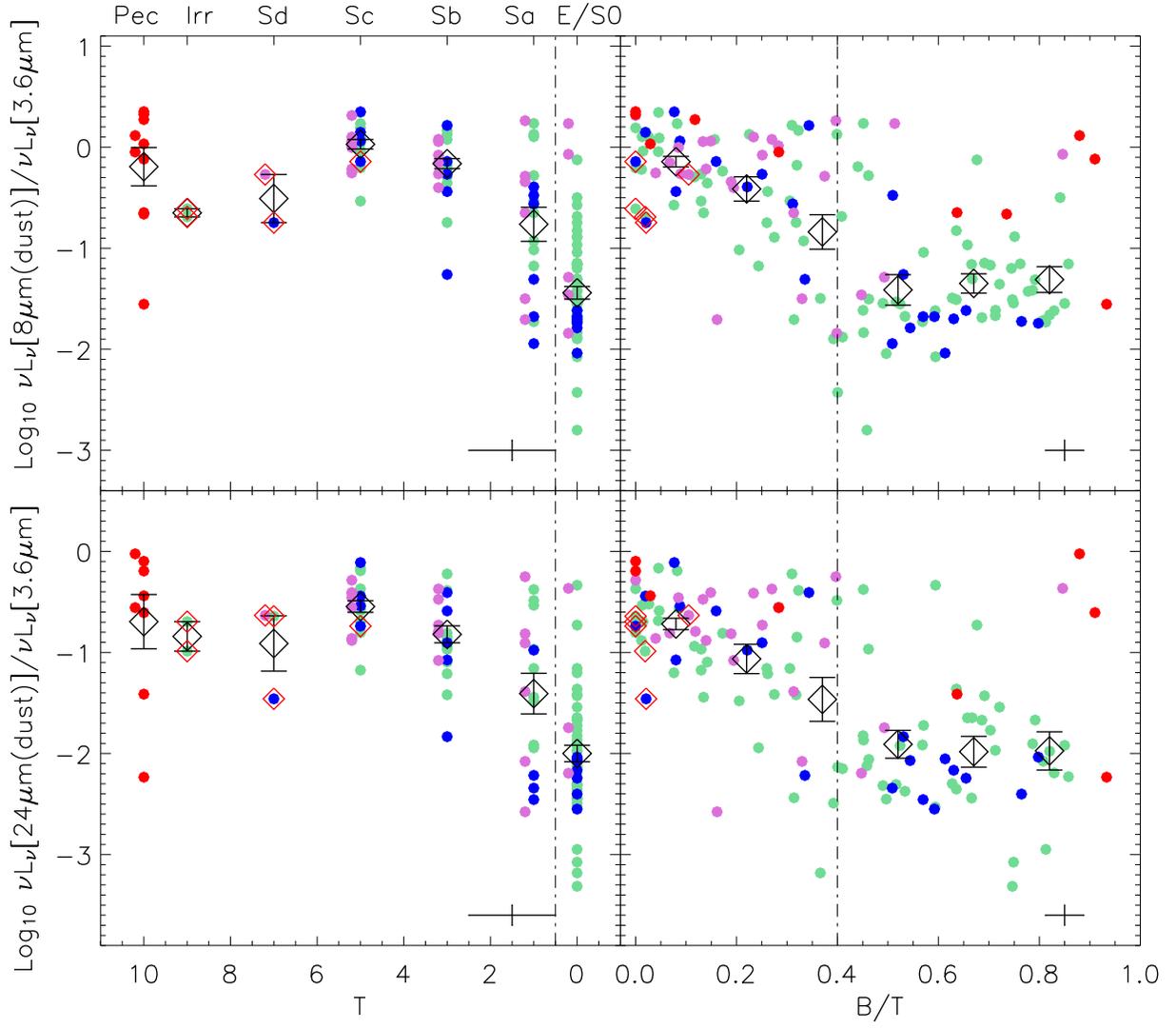}
\figcaption{Ratio of $8\mu m$ and $24\mu m$ dust-to-star ratios
    plotted against $T$ and $B/T$. Symbols are the same as defined in Figure~\ref{fig:T.Bratio}.
    The typical errors are plotted in the lower right corners.
\label{fig:mor.ratio} }
\end{figure}

The MIR dust-to-star ratios are also plotted against different
morphological types, in Figure~\ref{fig:mor.ratio}. The prominent
correlations are also detected between the MIR dust-to-star ratios
and both the Hubble $T$ types and the $B/T$ ratios. Furthermore,
such correlations seem to be more obvious than those between the MIR
luminosities and morphological types. Since both the $8\mu m$ and
$24\mu m$ dust luminosities possess correlations with SFRs for
normal galaxies, and the $3.6\mu m$ luminosity is a reliable tracer
of stellar component, such ratios can be treated as the dust-to-star
ratios. The correlations between the MIR dust-to-star ratios and the
morphological types could represent the distribution of SFRs per
unit stellar mass \citep{Wen2007} over different morphologies.
Contrary to the behavior in Figure~\ref{fig:mor.lum}, the dwarf
galaxies are roughly consistent with the other late-type galaxies
within the error bars, but still present a little lower dust-to-star
ratios. This could be explained with the fact that the gravitation
potentials of these low mass galaxies may not be strong enough to
retain as much dust and gas against the radiation fields as those of
the normal-mass galaxies, and hereby all the dwarf galaxies show
slightly lower MIR dust-to-star ratios. Almost all the peculiar
galaxies present higher MIR dust-to-star ratios.

\begin{figure}
\epsscale{1.1}
\plotone{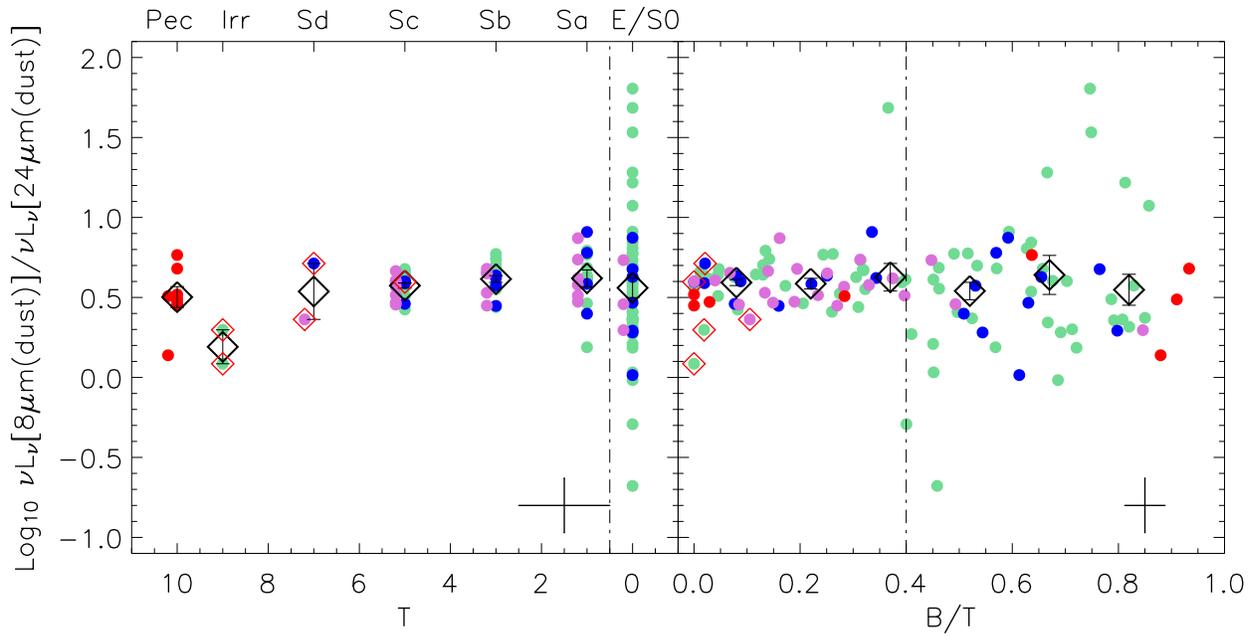}
\figcaption{Distribution of ratios between $8\mu m$ and $24\mu m$ dust luminosities
    subdivided with different morphological types.
    Symbols are defined the same as in Figure~\ref{fig:T.Bratio}.
    The typical errors are plotted in the lower right corners.
\label{fig:mor.dust} }
\end{figure}

The ratio of the $8\mu m$ dust luminosity to the $24\mu m$ dust
luminosity is compared with different morphological types in
Figure~\ref{fig:mor.dust}. Considering the $8\mu m$ dust emissions
mainly represent emission of PAHs heated by B type stars while
$24\mu m$ dust emissions stand for emissions from hot dust mostly
heated by O type stars \citep{Peeters2004}, the ratio can be treated
as the estimation of ratio between these two components. In general,
$8\mu m$(dust)-to-$24\mu m$(dust) ratios remain constant on average
along the Hubble sequence, while the larger scatter in distribution
of the early-type probably arises from existence of nuclear AGNs
which destroy PAH emissions presumably due to photodestruction of
the PAH molecules by EUV/X-ray photons
\citep{Genzel1998,Siebenmorgen2004} or the outer diffuse PAH
emissions heated by the older stars \citep{Sauvage2005}. This can
also be seen in later discussion in Figure~\ref{fig:SF.AGN} where
AGNs present lower $8\mu m$(dust) luminosities. It should be noted
that the two dwarf irregular galaxies exhibit lower $8\mu
m$(dust)-to-$24\mu m$(dust) ratios than the most of the late-type
do, possibly due to their low metallicities
\footnote{with metallicity $12+\log(O/H)=8.55$ and $8.35$
respectively, about $3/4$ and $1/2$ the solar value
\citep{Asplund2004}.}
\citep{Engelbracht2005}.

The above result indicates that for the normal galaxies except the
dwarfs, the ratio of these two dust components does not vary much
against morphology. Therefore the $8\mu m$(dust) luminosities are as
capable to trace the galaxy SFRs as the $24\mu m$(dust) luminosities
as \citet{Wu2005a} has pointed out.

\subsection{Statistics of MIR Properties}\label{subsec:statistics}
In order to compare the statistical MIR properties of different
types of sample galaxies, we performed all the following statistics
on the sub-sample of 125 galaxies which have photometric information
for all the optical and MIR bands. Since from
Figure~\ref{fig:properties} and K-S test results this sub-sample
does not exhibit notable differences from the 154-galaxy sample, we
suggest that the following statistical discussions are
representative of all the sample galaxies.

\begin{figure}[t]
\epsscale{1.0} \plotone{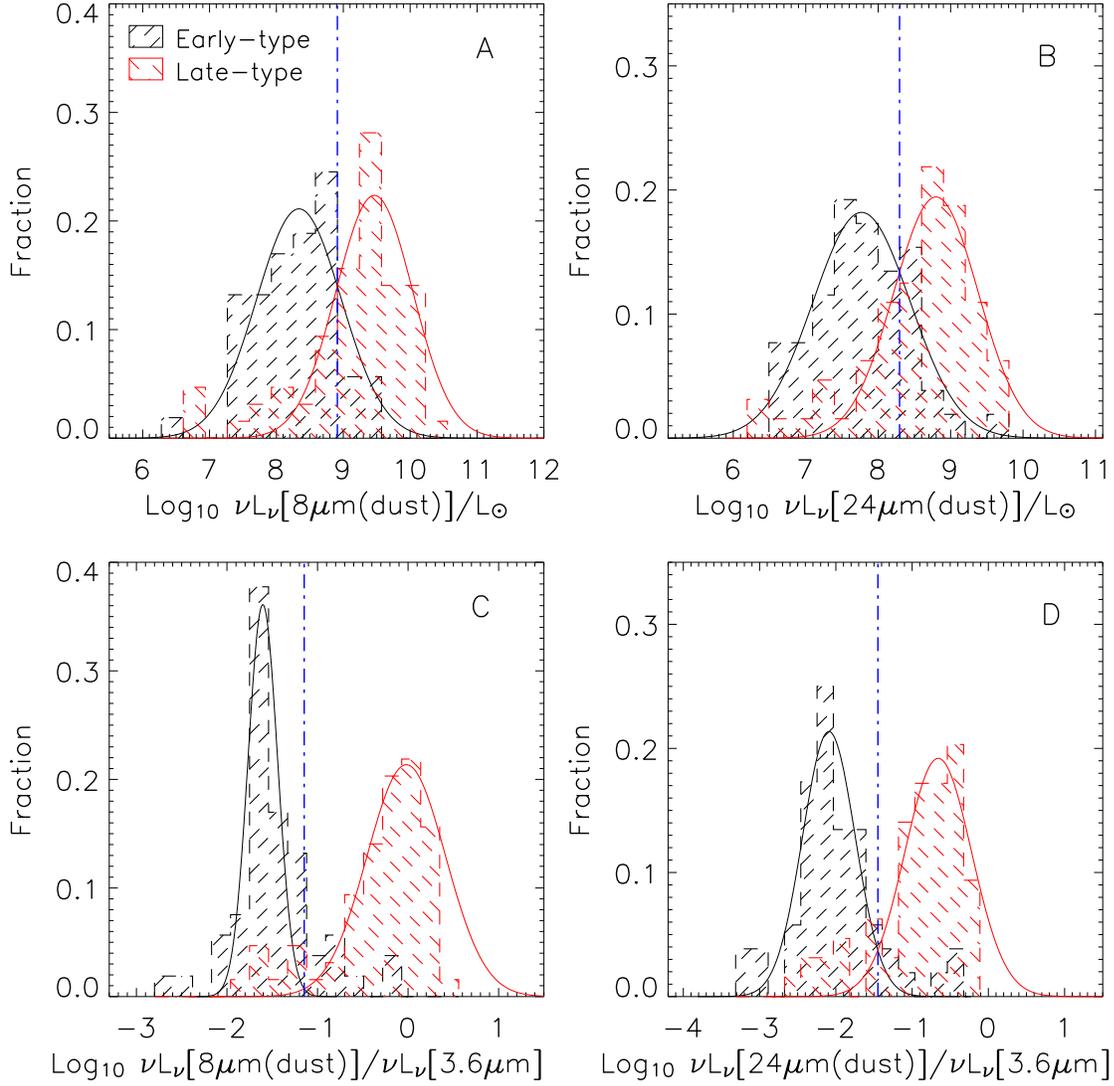} \figcaption{Distributions of MIR
luminosities and dust-to-star ratios for early-type
    and late-type galaxies of the 125 sub-sample.
    A: distribution of $8\mu m$(dust) luminosities;
    B: $24\mu m$(dust) luminosities;
    C: $8\mu m$ dust-to-star ratios;
    D: $24\mu m$ dust-to-star ratios.
    The black histogram represents distribution of early-types,
    and the red for late-types.
    The fitted Gaussian of each distribution is overlaid, with dividing lines
    through intersection points of Gaussian marked.
\label{fig:early.late}
}
\end{figure}

\begin{figure}[t]
\epsscale{1.0} \plotone{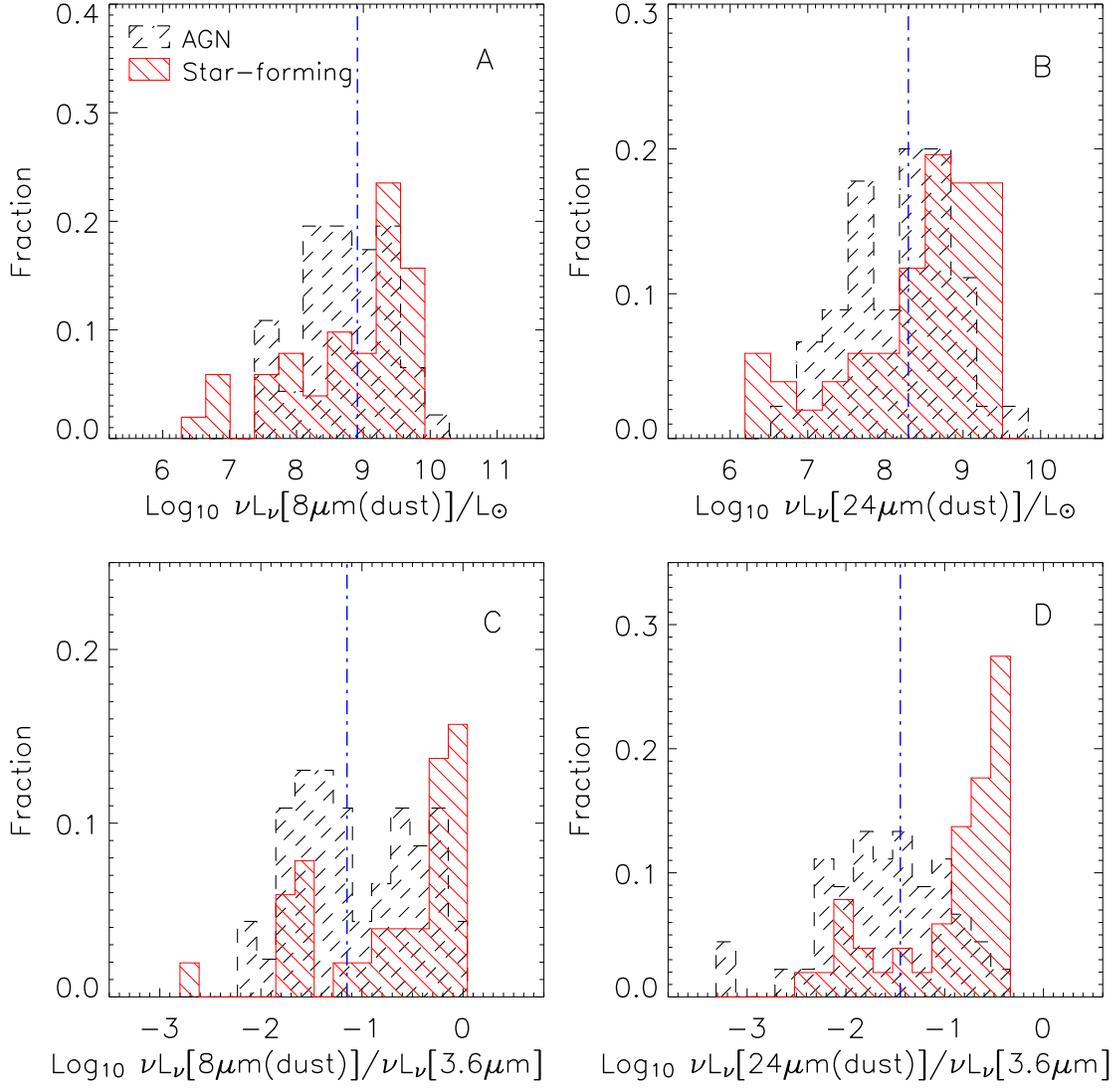} \figcaption{Distributions of MIR
luminosities and dust-to-star ratios in
    Star-forming galaxies and galaxies with AGN activities
    (as in Figure~\ref{fig:early.late}).
    The red histogram represents distribution of Star-forming galaxies,
    and the black for AGNs. Division lines are the same as in
    Figure~\ref{fig:early.late}.
\label{fig:SF.AGN} }
\end{figure}

Distributions of the MIR luminosities and dust-to-star ratios for
both the early-type and the late-type galaxies in the sub-sample are
presented in Figure~\ref{fig:early.late}. It is clear that, in
general, the late-type galaxies exhibit quite different MIR
properties to early-type ones, presenting a distinct and
statistically higher MIR luminosities and dust-to-star ratios. The
Gaussian fitting is performed on each distribution, and the lines
crossing intersection points of the two Gaussian distributions are
presented as the division of the two morphological types. In Part A
of Table~\ref{tab:type}, the median values and the mean values
together with scatters of distributions of the early-type and the
late-type galaxies are listed. Probabilities that the distributions
of these two types can match are all smaller than $8.8\times
10^{-5}$, indicating that they present quite distinct properties.
Therefore, both the MIR luminosities and dust-to-star ratios can be
used to separate the early-type galaxies from the late-type ones.
Table~\ref{tab:type.division} displays specific values of divisions
and the reliability of such classifications. We define the {\rm
reliability} of classification as the fraction of galaxies from the
subsample that are selected by Hubble $T$ types. For example, out of
the 117 normal galaxies which have images in all IRAC bands, the T
type criterion selects 64 late-type galaxies and 53 early-types. 47
of the 64 late-type galaxies have $\log\nu L_{\nu}(8\mu
m)/L_{\odot}\geqslant8.91$ and thus are consistently classified as
late-type, while 47 of the 53 early-types have $\log\nu L_{\nu}(8\mu
m)/L_{\odot}\leqslant8.91$ and thus are classified as early-type.
Therefore the $\log\nu L_{\nu}(8\mu m)/L_{\odot}=8.91$ maintains
reliability of 73\% for selecting late-type galaxies and 89\% for
early-types. Correspondingly, $\log\nu L_{\nu}(24\mu
m)/L_{\odot}=8.30$ gives a reliability of 70\% for the late-type and
77\% for the early-type; $\log\nu L_{\nu}(8\mu m)/\nu L_{\nu}(3.6\mu
m)=-1.15$ can give a 88\% for the late-type and 83\% for the
early-type; and $\log\nu L_{\nu}(24\mu m)/\nu L_{\nu}(3.6\mu
m)=-1.45$ can give a 84\% for the late-type and 85\% for the
early-type; Hereby the MIR dust luminosities especially the MIR
dust-to-star ratios can be effective tools to divide the early-type
galaxies from the late-type ones.

The comparisons between MIR properties of star-forming galaxies and
AGNs (see \S\ref{subsec:sample}) are shown in
Figure~\ref{fig:SF.AGN}. Generally, star-forming galaxies present
stronger MIR emissions than those possess AGN activities. From Part
B of Table~\ref{tab:type}, the star-forming galaxies and AGNs show
the probability of matching each other in $8\mu m$ and $24\mu m$
dust luminosites and dust-to-star ratios with less than $2.1\times
10^{-3}$, indicating quite different MIR properties, although not so
distinct as these between the early-type and late-type ones.

\begin{figure}[t]
\epsscale{1.0} \plotone{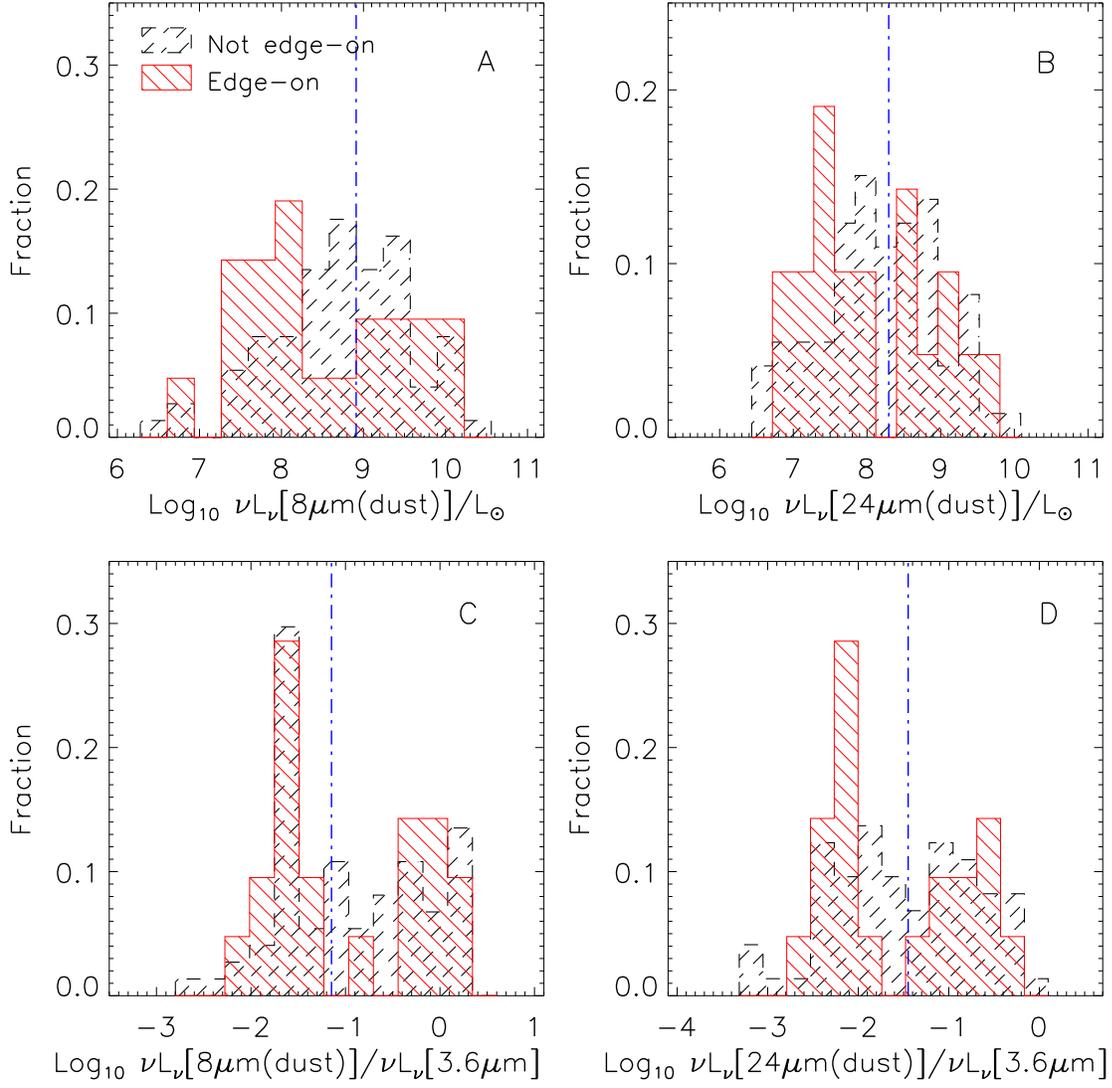} \figcaption{Distributions of MIR
luminosities and dust-to-star ratios of edge-on and not edge-on
galaxies (as in Figure~\ref{fig:early.late}).
    The black histogram represents distribution of not edge-on galaxies,
    and the red for the edge-on ones. Division lines are the same as in
    Figure~\ref{fig:early.late}.
\label{fig:edge.on} }
\end{figure}

\begin{figure}[t]
\epsscale{1.0} \plotone{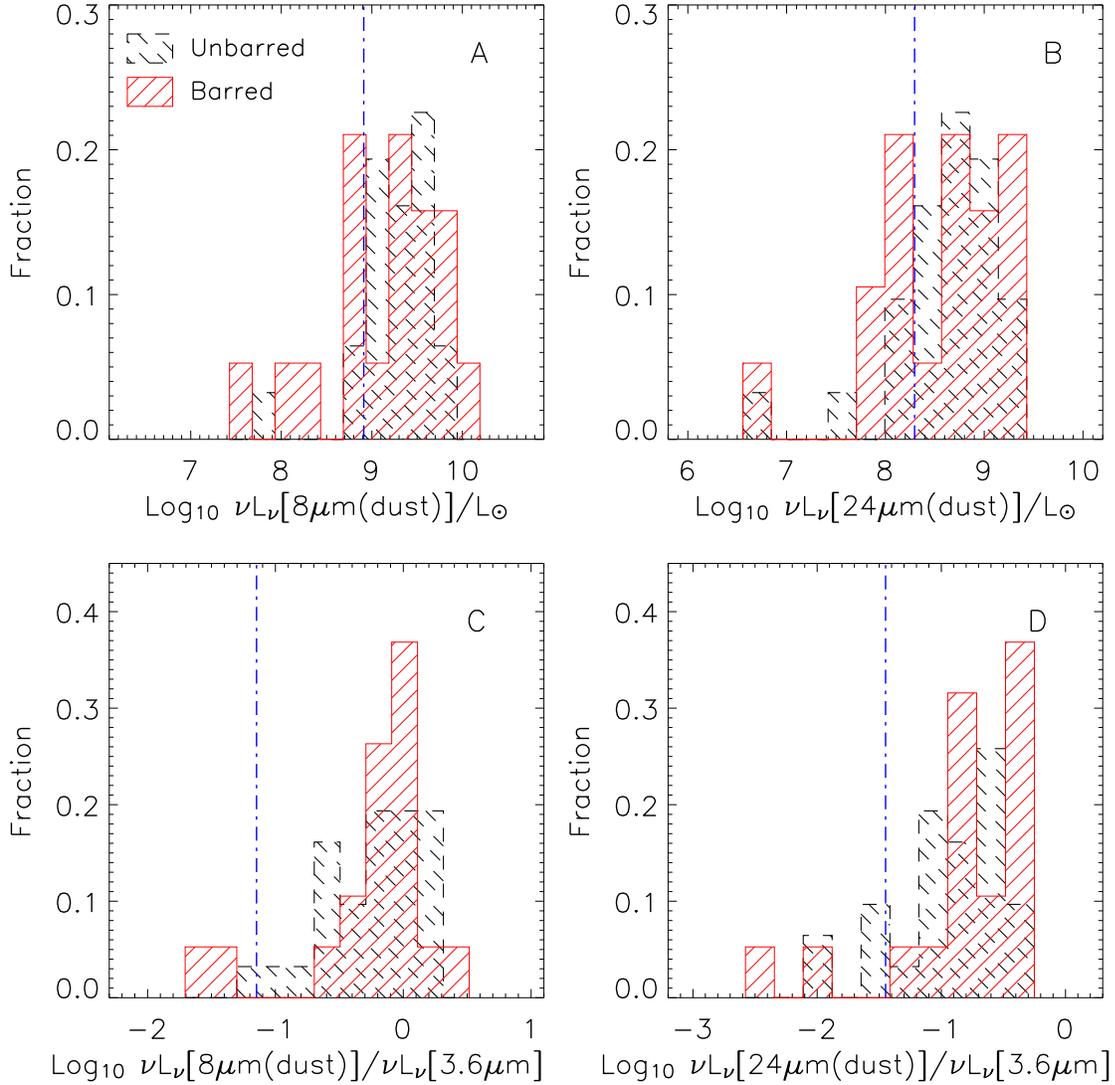} \figcaption{Distributions of MIR
luminosities and dust-to-star ratios in
    barred and unbarred galaxies (as in Figure~\ref{fig:early.late}).
    The red histogram represents distribution of barred spirals,
    and the black for the unbarred. Division lines are the same as in
    Figure~\ref{fig:early.late}.
\label{fig:bar} }
\end{figure}

Some of the sample galaxies are edge-on galaxies and some are barred
galaxies. Will the galaxies in edge-on view or with bars present
different MIR properties? We compare the distributions of the MIR
dust luminosities and the MIR dust-to-star ratios between edge-on
galaxies, barred galaxies and normal galaxies in
Figure~\ref{fig:edge.on} and Figure~\ref{fig:bar}. The statistical
results are listed in Part C and D of Table~\ref{tab:type}, with the
probabilities in matching each by few to tenth percent, indicating
rather resembling distributions. Thus, neither galaxies in edge-on
view nor galaxies with bars present statistical differences from the
other normal galaxies in both MIR properties. Furthermore, although
based on limited data points, one can still find out in
Figure~\ref{fig:edge.on} that the edge-on galaxies can also be
classified into early-type and late-type ones with the division
criteria described in Table~\ref{tab:type.division}. Therefore, the
inclination of galaxies and the existence of bars do not affect our
previous results.

We have also checked the distributions of total sample of 154
galaxies in the above parameters and type comparisons, and yielded
quite similar results, therefore the selection of this sub-sample
does not affect the reliability of our statistical results.

\section{SUMMARY}\label{section:summary}
We investigated the correlations between the morphological types and
the MIR properties of a local optically flux-limited sample of
galaxies selected from the spectroscopic catalogue of galaxies in
SDSS-DR4, cross-correlated with the Lockman Hole, ELAIS-N1 and
ELAIS-N2 of {\it Spitzer} SWIRE survey. Aperture photometry has been
performed on all these galaxies in all optical and MIR bands.
Morphological classifications have been performed by both visual
inspection and the bulge-disk decomposition with GIM2D. Our major
results are as follows:

{\noindent (1) The presented analysis clearly shows that the
bulge-to-total ratio $B/T$ obtained by the bulge-disk decomposition
is proved a qualified quantitative measure of the Hubble $T$ types.
Galaxies with earlier morphological types possess larger bulge
ratios while later type ones have more dominant disk structures. }

{\noindent (2) The $3.6\mu m$ luminosity presents a tight
correlation with the stellar mass, and this provides us a new tool
to estimate the stellar mass. The empirical formula to calculate the
stellar mass by the $3.6\mu m$ luminosity is given in
Equation~\ref{eqn:I1.mass}. }

{\noindent (3) Except for the dwarf galaxies and a few peculiar
objects, the MIR dust luminosities of $8\mu m$ and $24\mu m$ exhibit
correlations with either Hubble $T$ type or the bulge-to-total ratio
$B/T$. Such correlations are much more obvious if we used the MIR
dust-to-star ratios (either $\nu L_{\nu}[8\mu m (dust)]/\nu
L_{\nu}[3.6\mu m]$ or $\nu L_{\nu}[24\mu m (dust)]/\nu
L_{\nu}[3.6\mu m]$) instead of the MIR luminosities. }

{\noindent (4) The MIR dust luminosity ratios of $\nu L_{\nu}[8\mu m
(dust)]/\nu L_{\nu}[24\mu m(dust)]$ turn to be roughly constant
against the morphological types, especially for the late-type
galaxies. Therefore, on average, the $8\mu m$(dust) luminosity can
as reliably measure the global SFRs of normal galaxies as the $24\mu
m$(dust) luminosity can, regardless of the morphological types. }

{\noindent (5) Distributions of the MIR dust luminosities and the
MIR dust-to-star ratios of both the early-type and the late-type
galaxies are very different. The late-type galaxies present higher
MIR dust luminosities and MIR dust-to-star ratios than the
early-type galaxies. The MIR dust luminosities of $8\mu m$ and
$24\mu m$ and especially the MIR dust-to-star ratios of $\nu
L_{\nu}[8\mu m (dust)]/\nu L_{\nu}[3.6\mu m]$ and $\nu L_{\nu}[24\mu
m (dust)]/\nu L_{\nu}[3.6\mu m]$ can provide an effective tool to
distinguish the late-type galaxies from the early-types. }

{\noindent (6) The star-forming galaxies and AGNs also present
different statistical MIR properties, with the former showing higher
MIR luminosities and MIR dust-to-star ratios. The statistical
results show that either galaxies in edge-on view or galaxies with
bars do not present quite different MIR properties from other sample
galaxies. }

\acknowledgements

The authors would like to appreciate the anonymous referee for
helpful comments and suggestions, and we are grateful for the help
of D.B. Sanders, S. Mao, Z.-L. Zhou, X.-Y. Xia, Z.-G. Deng, Z. Wang,
J.-S. Huang, C.-N. Hao, F.-S. Liu and J.-L. Wang. This project was
supported by the National Science Foundation of China through grants
10273012, 10333060 and 10473013.

This work is based in part on observations made with the Spitzer
Space Telescope, which is operated by the Jet Propulsion Laboratory,
California Institute of Technology under NASA contract 1407.

Funding for the SDSS and SDSS-II has been provided by the Alfred P.
Sloan Foundation, the Participating Institutions, the National
Science Foundation, the U.S. Department of Energy, the National
Aeronautics and Space Administration, the Japanese Monbukagakusho,
the Max Planck Society, and the Higher Education Funding Council for
England. The SDSS is managed by the Astrophysical Research
Consortium for the Participating Institutions. The Participating
Institutions are the American Museum of Natural History,
Astrophysical Institute Potsdam, University of Basel, Cambridge
University, Case Western Reserve University, University of Chicago,
Drexel University, Fermilab, the Institute for Advanced Study, the
Japan Participation Group, Johns Hopkins University, the Joint
Institute for Nuclear Astrophysics, the Kavli Institute for Particle
Astrophysics and Cosmology, the Korean Scientist Group, the Chinese
Academy of Sciences (LAMOST), Los Alamos National Laboratory, the
Max-Planck-Institute for Astronomy (MPIA), the Max-Planck-Institute
for Astrophysics (MPA), New Mexico State University, Ohio State
University, University of Pittsburgh, University of Portsmouth,
Princeton University, the United States Naval Observatory, and the
University of Washington.

\clearpage

\clearpage

\clearpage

\clearpage

\clearpage

\clearpage

\clearpage

\clearpage

\clearpage

\clearpage

\clearpage

\clearpage

\clearpage

\clearpage

\begin{deluxetable}{lccccccc|c}
\tablecolumns{9}
\tabletypesize{\footnotesize}
\tablewidth{0pt}
\tablecaption{The numbers of different morphological types of sample
galaxies\label{tab:mor.class} } \tablehead{
    \colhead{} &
    \colhead{E/S0} &
    \colhead{Sa} &
    \colhead{Sb} &
    \colhead{Sc} &
    \colhead{Sd} &
    \colhead{Irr} &
    \colhead{Normal} &
    \colhead{Pec}
}
\startdata
 General   &    53    &    13   &    13   &    11   &    1    &    2   &    93   &  7  \\
 Barred    &    6     &    6    &    8    &    7    &    1    &    0   &    28   &  2  \\
 Edge-on   &    7     &    6    &    5    &    4    &    1    &    0   &    23   &  1  \\
 Total     &    66    &    25   &    26   &    22   &    3    &    2   &   144   &  10 \\
Fraction\% &   42.9   &   16.2  &   16.9  &   14.3  &   1.9   &   1.3  &   93.5  & 6.5 \\
\enddata
\tablecomments{Respective numbers and fractions of different morphological
types and classes of 154 sample galaxies.
Note that throughout the discussion in this paper,
unless particularly stated, general, barred and edge-on types are referred
to normal galaxies (from T=0 to T=9), while peculiar as an independent class.}
\end{deluxetable}

\clearpage

\begin{deluxetable}{l|cccccccc}
\tablecolumns{9}
\tabletypesize{\footnotesize}
\tablewidth{0pt}
\tablecaption{The numbers of different morphological types observed
by each {\it Spitzer} MIR band\label{tab:mor.spitzer} } \tablehead{
    \colhead{Band} &
    \colhead{E/S0} &
    \colhead{Sa} &
    \colhead{Sb} &
    \colhead{Sc} &
    \colhead{Sd} &
    \colhead{Irr} &
    \colhead{Pec} &
    \colhead{Total}}
\startdata
3.6$\mu$m  &  61   &  22   &  26   &  21   &  3   &  2   &  10   &  145  \\
4.5$\mu$m  &  66   &  24   &  25   &  21   &  2   &  2   &  10   &  150  \\
5.8$\mu$m  &  61   &  22   &  26   &  21   &  3   &  2   &  10   &  145  \\
8.0$\mu$m  &  66   &  24   &  25   &  21   &  2   &  2   &  10   &  150  \\
24$\mu$m   &  57   &  22   &  23   &  22   &  3   &  2   &   8   &  137  \\
all bands  &  53   &  19   &  20   &  21   &  2   &  2   &   8   &  125  \\
\enddata
\tablecomments{The list shows respective numbers of different
morphological types in each of the four IRAC bands and MIPS 24$\mu$m
band, and also in the sub-sample whose 125 galaxies have images in
all MIR bands.}
\end{deluxetable}

\clearpage

\begin{deluxetable}{lcccccccc}
\tablecolumns{9}
\tabletypesize{\footnotesize}
\tablewidth{0pt}
\rotate
\tablecaption{MIR Properties Distribution and Statistics\label{tab:type}
}
\tablehead{
    \colhead{Parameters} &
    \multicolumn{2}{c}{Median} &
    \multicolumn{2}{c}{Mean(scatter)} &
    \multicolumn{2}{c}{Number} &
    \colhead{D} &
    \colhead{P}\\
\colhead{(1)} & \multicolumn{2}{c}{(2)} & \multicolumn{2}{c}{(3)} & \multicolumn{2}{c}{(4)} & \colhead{(5)} & \colhead{(6)}
}
\startdata
$<Part A>$ & \colhead{$Early$} & \colhead{$Late$} & \colhead{$Early$} & \colhead{$Late$}
& \colhead{$Early$} & \colhead{$Late$} & \colhead{} & \colhead{} \\
\hline
$\log \nu L_{\nu}(8\mu m)/L_{\odot}$ & 8.30 & 9.30 &  8.28(0.09) & 9.13(0.10) & 53 & 64 & 0.72 & $2.9\times10^{-6}$\\
$\log \nu L_{\nu}(24\mu m)/L_{\odot}$ & 7.76 & 8.67 & 7.77(0.09) & 8.55(0.10) & 53 & 64 & 0.63 & $8.8\times10^{-5}$ \\
$\log \nu L_{\nu}(8\mu m)/\nu L_{\nu}(3.6\mu m)$ & -1.55 &-0.22 & -1.49(0.07)&-0.33(0.07)&53&64&0.83& $4.7\times10^{-8}$ \\
$\log \nu L_{\nu}(24\mu m)/\nu L_{\nu}(3.6\mu m)$&-2.06&-0.81&-2.00(0.08)&-0.92(0.07)&53&64&0.77&$6.9\times10^{-7}$ \\
\hline
$<Part B>$ & \colhead{\it Star-forming} & \colhead{$AGN$} & \colhead{\it Star-forming} & \colhead{$AGN$}
& \colhead{\it Star-forming} & \colhead{$AGN$} & \colhead{} & \colhead{} \\
\hline
$\log \nu L_{\nu}(8\mu m)/L_{\odot}$ &  9.31 & 8.78  &  8.99(0.14)  &  8.72(0.10)  & 51 & 46 & 0.38 & $1.8\times10^{-3}$ \\
$\log \nu L_{\nu}(24\mu m)/L_{\odot}$ &  8.75 &  8.27  &  8.49(0.13) &  8.16(0.10) & 51 & 46 & 0.38 & $2.1\times10^{-3}$ \\
$\log \nu L_{\nu}(8\mu m)/\nu L_{\nu}(3.6\mu m)$ & -0.13 & -1.16 & -0.37(0.10) & -1.04(0.09) & 51 & 46 & 0.38 & $1.8\times10^{-3}$\\
$\log \nu L_{\nu}(24\mu m)/\nu L_{\nu}(3.6\mu m)$ & -0.70 & -1.65 & -0.87(0.09)& -1.60(0.10) & 51 & 46 & 0.38 & $2.1\times10^{-3}$ \\
\hline $<Part C>$ & \colhead{\it Edge-on} & \colhead{$Not$} &
\colhead{\it Edge-on} & \colhead{$Not$}
& \colhead{\it Edge-on} & \colhead{$Not$} & \colhead{} & \colhead{} \\
\hline
$\log \nu L_{\nu}(8\mu m)/L_{\odot}$ & 8.14 & 8.81 & 8.52(0.22) & 8.80(0.08) & 21 & 96 &  0.33 & 0.05 \\
$\log \nu L_{\nu}(24\mu m)/L_{\odot}$ & 7.82 & 8.31 & 7.96(0.22) & 8.21(0.09) & 21 & 96 &  0.31  & 0.08  \\
$\log \nu L_{\nu}(8\mu m)/\nu L_{\nu}(3.6\mu m)$ & -1.26 & -0.75 & -0.94(0.19) & -0.83(0.08) & 21 & 96 & 0.25 & 0.25 \\
$\log \nu L_{\nu}(24\mu m)/\nu L_{\nu}(3.6\mu m)$ & -1.83 & -1.37 & -1.51(0.18) & -1.42(0.09) & 21 & 96 & 0.22 & 0.40 \\
\hline $<Part D>$ & \colhead{$Barred$} & \colhead{$Unbarred$} &
\colhead{$Barred$} & \colhead{$Unbarred$}
& \colhead{$Barred$} & \colhead{$Unbarred$} & \colhead{} & \colhead{} \\
\hline
$\log \nu L_{\nu}(8\mu m)/L_{\odot}$ & 9.30 & 9.44 & 9.12(0.16) & 9.26(0.15) & 19 & 31 & 0.24 & 0.51 \\
$\log \nu L_{\nu}(24\mu m)/L_{\odot}$ & 8.66 & 8.78 & 8.54(0.16) & 8.67(0.14) & 19 & 31 & 0.17 & 0.86 \\
$\log \nu L_{\nu}(8\mu m)/\nu L_{\nu}(3.6\mu m)$ & -0.15 & -0.22 & -0.26(0.12) & -0.27(0.09) & 19 & 31 & 0.16 & 0.91 \\
$\log \nu L_{\nu}(24\mu m)/\nu L_{\nu}(3.6\mu m)$ & -0.73 & -0.81 & -0.84(0.14) & -0.86(0.08) & 19 & 31 & 0.21 & 0.68 \\
\enddata
\tablecomments{Col. (2) are median values of the two compared types
of sample galaxies. Col.(3) are mean values with scatters. Col.(4)
are the respective numbers in statistics. Col.(5) are K-S test
D(discrepancy) with Col.(6) the P(probability) that the two
distributions match. The four parts correspond to: Part A are
early-type and late-type sample galaxies; Part B are Star-forming
and AGN samples; Part C are edge-on and not edge-on samples; Part D
are are barred and unbarred spirals.}
\end{deluxetable}

\clearpage

\begin{deluxetable}{lccc}
\tablecolumns{4}
\tabletypesize{\footnotesize}
\tablewidth{0pt}
\tablecaption{Divisions of Early-type and Late-type Galaxies \label{tab:type.division}
}
\tablehead{
    \colhead{Parameters} &
    \colhead{Divisions} &
    \colhead{Reliability(L)} &
    \colhead{Reliability(E)}\\
\colhead{(1)} & \colhead{(2)} & \colhead{(3)} & \colhead{(4)}
}
\startdata
$\log \nu L_{\nu}(8\mu m)/L_{\odot}$ & 8.91 & 73\% & 89\% \\
$\log \nu L_{\nu}(24\mu m)/L_{\odot}$ & 8.30 & 70\% & 77\% \\
$\log \nu L_{\nu}(8\mu m)/\nu L_{\nu}(3.6\mu m)$ & -1.15 & 88\% & 83\% \\
$\log \nu L_{\nu}(24\mu m)/\nu L_{\nu}(3.6\mu m)$ & -1.45 & 84\% & 85\% \\
\enddata
\tablecomments{Col.(2) are the values corresponding to the division of the two types.
Col.(3) are the reliability for selecting late-type galaxies,
while Col.(4) are the reliability for selecting early-types.}
\end{deluxetable}

\clearpage

\end{document}